\documentclass[aps,pra,reprint,superscriptaddress]{revtex4-2}
\usepackage{amsmath}
\usepackage{amsfonts}
\usepackage{amssymb}
\usepackage{graphicx}
\usepackage{dcolumn}
\usepackage{bm}
\usepackage{makecell}
\usepackage{multirow}
\usepackage{hyperref}
\usepackage{setspace}
\hypersetup{
	colorlinks=true,
	linkcolor=black,
	citecolor=black,
	urlcolor=black
}

\begin{document}
\title{Aging-Improved NH$_3$ Sensing in Polypyrrole Nanotube Chemiresistors}
\author{Arya Mohan}
\email{arya.mohan@uclouvain.be}
\affiliation{Nanoscopic Physics (NAPS), Institute of Condensed Matter and Nanosciences (IMCN), Université catholique de Louvain (UCLouvain), 1348 Louvain-la-Neuve, Belgium}

\author{Olivia Kamdom}
\affiliation{Nanoscopic Physics (NAPS), Institute of Condensed Matter and Nanosciences (IMCN), Université catholique de Louvain (UCLouvain), 1348 Louvain-la-Neuve, Belgium}

\author{Antoine Enel}
\affiliation{Institute of Information and Communication Technologies, Electronics and Applied Mathematics (ICTEAM), Université catholique de Louvain (UCLouvain), 1348 Louvain-la-Neuve, Belgium}
\affiliation{VOCSens, 6 Chemin du Cyclotron, 1348 Louvain-la-Neuve, Belgium}

\author{Beno\^{i}t Hackens}
\email{benoit.hackens@uclouvain.be}
\affiliation{Nanoscopic Physics (NAPS), Institute of Condensed Matter and Nanosciences (IMCN), Université catholique de Louvain (UCLouvain), 1348 Louvain-la-Neuve, Belgium}

\begin{abstract}
	Polypyrrole NH$_3$ chemiresistors in nanotube (NT) and planar film (PF) morphologies are compared on response magnitude, limit of detection (LOD), sensitivity, and response/recovery kinetics, including the performance of the same sensors tracked over 60 days of ambient storage and repeated NH$_{3}$ exposure. In the case of fresh sensors, the time constants for the fast process ($\tau_1$) is shorter in the NT than in the PF at every concentration. NT is 38 \% faster on average than PF. The slow process ($\tau_2$) overlaps between two morphologies and shows no concentration  dependence. The  two time constants differentiate the two morphologies through aging: NT $\tau_1$ and $\tau_2$ decrease substantially and monotonically over 60 days, while the time constants of PF show no consistent trend. All measured sensing parameters, LOD, sensor response and sensitivity improve or at least remain stable with aging in the case of NT, while all parameters degrade in the PF case. These improved sensor performance and aging identify the nanotube morphology as the more robust and manufacturable choice for long-term NH$_3$ sensing applications.

\end{abstract}

\maketitle

\section*{Introduction}
 A sensor that has to be re-calibrated or replaced within a week of fabrication is not viable compared to one that is stable, or even self-improving, over a comparable period, regardless of which has the higher response on day one. Yet most reported comparisons of chemiresistive gas sensors characterize the sensors when freshly fabricated, leaving how the response, sensitivity and limit of detection (LOD) evolve over weeks or months largely unexplored. This challenges the long-term deployment of sensors in real-world applications. It also highlights the sensitivity–stability trade-off in chemiresistors, where higher initial sensitivity often comes at the expense of long-term stability, and vice versa \cite{Weng2025,Dong2025}.

\par This trade-off is particularly relevant for conducting polymer chemiresistors, an attractive platform for room-temperature gas sensing. They avoid the high operating temperatures required by many metal-oxide semiconductors and offer facile, low-cost processing \cite{Hunter2020, Lakard2015, wong2020}. Among these, polypyrrole (PPy) is widely used for ammonia (NH$_3$) detection, where exposure to the reducing gas increases the resistance of the $p$-type semiconducting polymer \cite{Carquigny2009,Joshi2011}. A substantial amount of work has focused on improving sensor performance (response, sensitivity, LOD). Strategies such as nanostructuring \cite{Hien2019, Stejskal2018,  Kwon2010,Setka2017,Yang2010}, templated growth\cite{She2021,Li2023, Kwon2012} and morphological control have been adopted to increase the surface-to-volume ratio and shorten diffusion paths compared with continuous films. In particular, PPy nanotubes synthesized with methyl orange (MO) templates have a well-defined nanotubular morphology, high aspect ratio, high conductivity and environmental stability \cite{Irina2017,Kopecky2017, Varga2015}. Whether these gains survive over time, however, has received comparatively little attention.

\par Aging  and conductivity degradation in PPy has been extensively investigated, but not much on a sensing device \cite{Kayanak2001, varga2017,Sixou1996,Amini2022}. The electrical conductivity of MO templated PPy nanotubes and template-free granular PPy  has been studied over three years of aging and under accelerated aging and was found to be decaying via deprotonation and backbone oxidation. This study found that PPy nanotubes degrades nearly two orders of magnitude more slowly than the non templated PPy, attributed to the residual MO and presence of sulfonate (SO$_{3}^{-}$) group which is known to increase stability in conducting polymers \cite{varga2017}. 
\par In this work,  two types of samples are examined template-free polypyrrole deposited as a planar film (PF) and PPy nanotubes (NT) with MO templates. Both are compared on ammonia (NH$_{3}$) sensing performance metrics and how those metrics change over 60 days of ambient storage and use. The NH$_3$ response and recovery kinetics are analyzed in terms of time constants and are compared as they evolve with aging. We find that the NT morphology presents not only faster response with a shorter time constant, but an improvement in all sensing parameters as it ages.
\par 
\section*{Results and Discussions}
\begin{figure*}[htbp]
	\centering
	\includegraphics[width=1\textwidth]{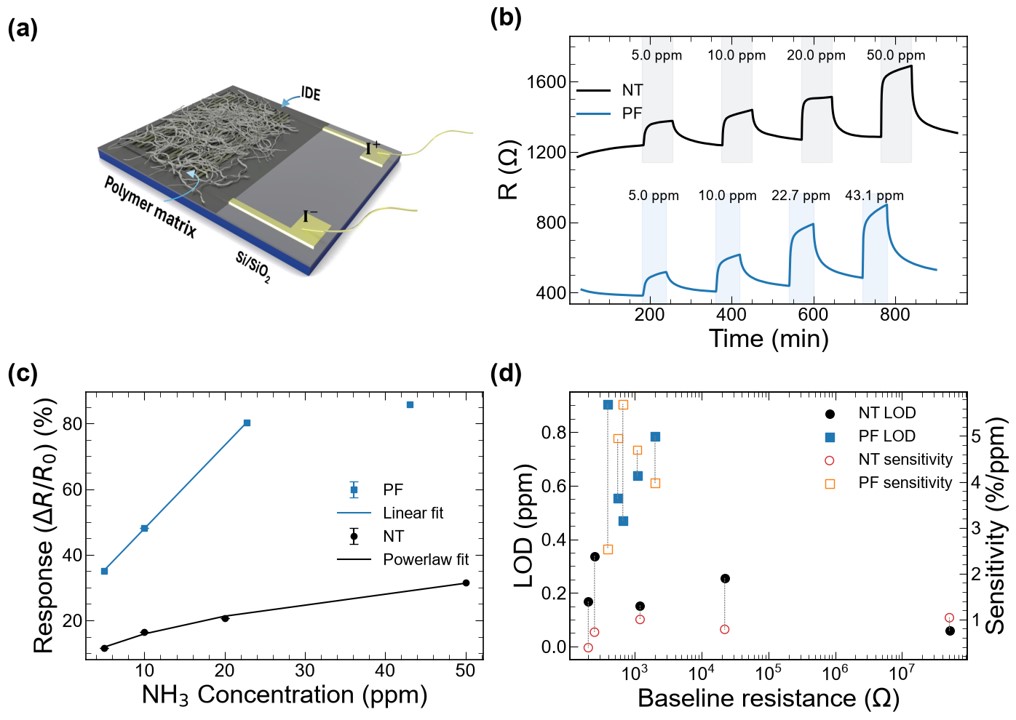}
	\caption{(a) Schematic of the polypyrrole (PPy) gas sensor device. (b) Dynamic resistance response of PPy nanotube (NT) sensors exposed to 5, 10, 20, and 50 ppm NH$_3$ for 75 min with a 120 min recovery period (indicated by the grey shaded region), alongside raw resistance data for PPy planar film (PF) exposed to 5, 10, 22, and 43 ppm NH$_3$ with a 60 min exposure and 120 min recovery (indicated by blue shaded region). (c) Sensor ($\Delta R/R_0$) \% as function of NH$_3$ concentration, with a linear fit for PF ($ < 40 \text{ ppm}$) and a power-law fit for NT. (d) Limit of detection (LOD) for PPy NT and PPy PF sensors as a function of baseline resistance; open symbols indicate the sensitivity of NT (at 5 ppm) and PF sensors.}
	\label{fig:1}
\end{figure*}
\begin{figure*}[htbp]
	\centering
	\includegraphics[width=1\textwidth]{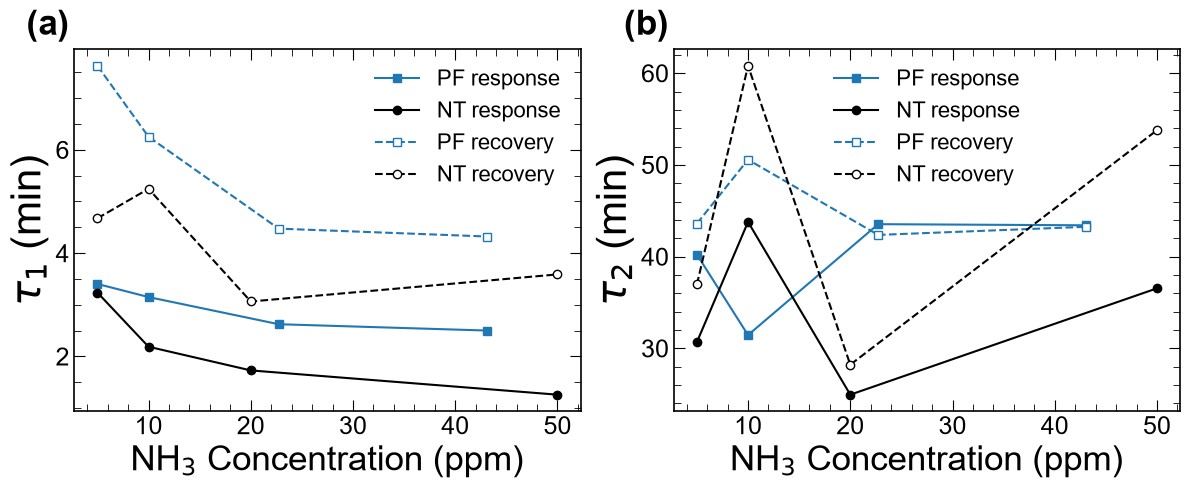}
	\caption{Response and recovery time constants ($\tau_1, \tau_2$) as a function of $\text{NH}_3$ concentration, obtained from double-exponential fits for PPy nanotube (NT) and planar film (PF) (closed and open symbols denote response and recovery times, respectively).}
	\label{fig:2}
\end{figure*}
The polypyrrole NT and PF preparation methods are described in the Materials and Methods section below and electron micrographs are shown in supplementary material. Following materials synthesis on interdigitated electrodes (IDEs) deposited on top of a piece of Si/SiO$_2$ substrate as shown schematically on Fig.\ref{fig:1}(a), we measured their resistance in controlled atmosphere, exposing them to specific concentrations of NH$_3$. Figure \ref{fig:1}(b) shows the resistance vs time of the PPy NT  and PPy PF under a sequential 5-50 ppm NH$_3$ exposure. Both sensors show an increase in resistance on exposure, which is typical for a $p$-type semiconductor like polypyrrole interacting with a reducing gas. A first look at the resistance change upon exposure allows to conclude that the saturation value is reached faster in PPy NT, compared with PF. Much faster saturation in PPy NT than PF indicates shorter time constants for the underlying process of gas adsorption and charge transfer. 
 
\par Response ($\Delta R/R_0$ \%, where $R_0$ is the baseline resistance) of both samples increases with NH$_3$ concentration as shown in Fig.\ref{fig:1}(c). The  PPy PF sensor has higher response of  35 \% at 5 ppm increasing to 86 \% at 43 ppm of NH$_3$, whereas PPy NT sample has a response of 11 \% at 5 ppm and 31 \% at 50 ppm of NH$_3$. The response depends linearly on concentration for PPy PF in the tested range of 5-22 ppm and tend to saturate at 43 ppm. It is fitted to straight line ($\Delta R/R_0 \propto C$) with a slope of 2.54 \%/ppm which is the sensitivity of PPy PF.  In contrast the response vs concentration of PPy NT does not show a linear trend and it is well fitted to a power law ($\Delta R/R_0 \propto C^n$ ) with exponent $n$  = 0.42. 
 \par Limit of detection (LOD) and sensitivity of various PPy NT and PF samples are compared based on their baseline resistance as shown in Fig.\ref{fig:1}(d). The LOD is determined using the standard signal-to-noise ratio criterion ($\text{LOD} = 3\sigma / \text{sensitivity}$), where $\sigma$ is the standard deviation of the baseline resistance noise. For PPy PF showing linear sensing behavior ($S = mC + C_0$), sensitivity is the slope ($m$). In contrast, PPy NT sensors exhibits a power-law response ($S = k C^n$), giving a concentration dependent sensitivity defined by the local slope $\frac{dS}{dC} = n k C^{n-1}$. The sensitivity for PPy NT is evaluated at 5 ppm for comparison.  The LOD and sensitivity for all PPy NT samples are lower than those of PPy PF. The PPy NT sensors have baseline resistances ranging from $200\ \Omega$ to $50\ \text{M}\Omega$, with LOD varying from $59$ to $337$ ppb across different devices. The PF samples have baseline resistance from $ 400\ \Omega $ to $2\ \text{k}\Omega$, with LOD in the range of $0.5$ to $0.9$ ppm. 
 \par The sensitivity varies from 0.9 to 1 \%/ppm for NT at 5 ppm and 2.5 to 5.6 \%/ppm sensors for PPy PF. Overall PPy NT sensor exhibit lower LOD, though its response and sensitivity is much lower compared to PPy PF. 
LOD values were obtained from each morphology's best-fit response model (power-law for NT, linear for PF). The detection limits of NT, in the tens of ppb range, are consistent with other reported PPy-based ammonia sensors \cite{Li2023,Gai2023,Yan2016} and are within the sensitivity requirements of applications such as breath analysis, environmental monitoring, and industrial leak detection \cite{Ricci2021, Lefferts2022, Aarya2020}.

\par Next, we focus on sensing dynamics and examine the transient response and recovery phases at each NH$_3$ concentration which were well fitted to a double exponential with two relaxation time constants $\tau_1$ (fast) and $\tau_2$  (slow) for every sensor (see supplementary material for a representative fit to the data). The fitting parameters are given in Tables S1 and S2. The two time constants,  $\tau_1$ in the range of few minutes and $\tau_2$ in the range of 20-60 min are shown in Fig.\ref{fig:2}.  The faster $\tau_1$ is decreasing with increasing NH$_3$ concentration both in response and recovery data for both PPy NT and PF. Whereas the slower $\tau_2$ is independent of NH$_3$ concentration. The time constant $\tau_1$ is shorter in PPy NT compared to PPy PF at every concentration in both response and recovery. On an average PPy NT response $\tau_1$ is 39 \% faster and recovery $\tau_1$ is 37 \% faster than PPy PF. In contrast, the mean response $\tau_2$ is shorter for PPy NT and mean recovery $\tau_2$ is identical, but the two distributions overlap for both morphologies. 

\begin{figure*}[htbp]
	\centering
	\includegraphics[width=1\textwidth]{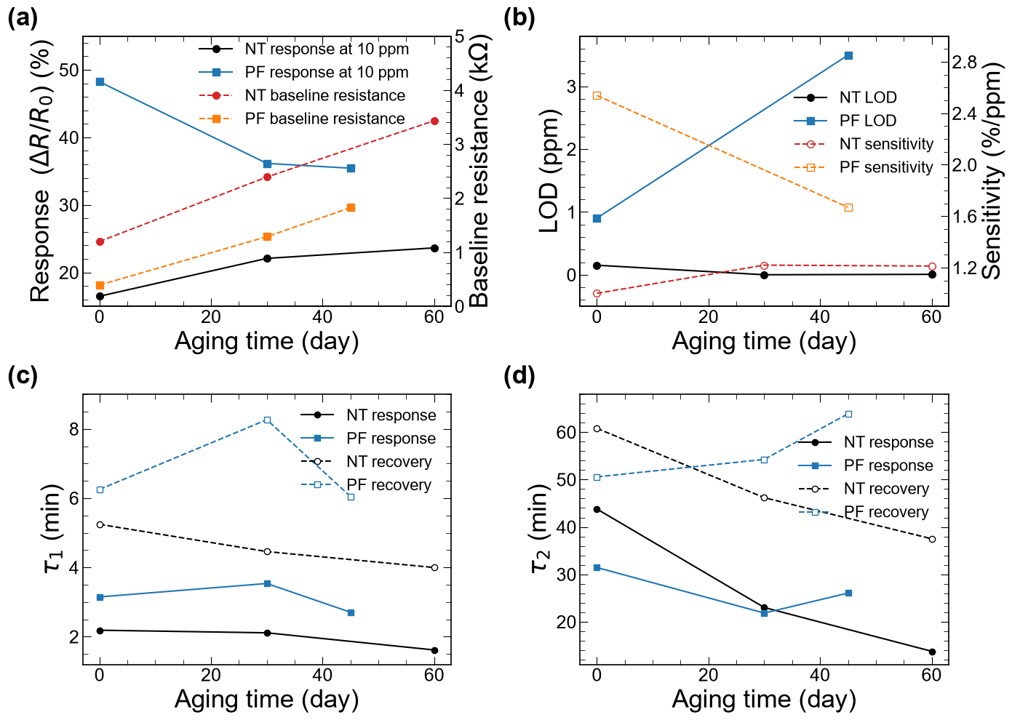}
	\caption{Aging characteristics of PPy sensors (a) Response ($\Delta R / R_0\%$) at $10\text{ ppm }\text{NH}_3$ and baseline resistance as a function of aging time for PPy nanotube (NT) and planar film (PF) sensors. (b) Limit of detection (LOD) as a function of aging time, alongside the sensitivity of NT ($5\text{ ppm}$) and PF sensors.(c, d) Response time constants ($\tau_1, \tau_2$) and recovery time constants as a function of aging time for both sensor types.}
	\label{fig:3}
\end{figure*}
\par The main result of this work concerns the influence of aging on gas sensing. To study this parameter, the resistance of all sensors are measured over 60 days after ambient storage and NH$_3$ exposure, as shown in Fig.\ref{fig:3}. The response ($\Delta R/R_0$) is increasing from 16.5 \% for 10 ppm for fresh sample to 23.6 \% after 60 days for PPy NT samples. However the baseline resistance is also increasing from 1.2 to 3.4 $\text{k}\Omega$.  The trend for response of PPy PF is opposite: it decreases from 48.23 to 35.4 \% after 45 days. The baseline resistance increases also in PPy PF from 392 $\Omega$ to 1.8 $\text{k}\Omega$. Increasing response of PPy NT as it ages has never been reported. In both morphologies the baseline resistance drifts substantially and more severely in PPy PF. This is direct evidence that conducting network itself is physically degrading with the age.  The LOD of PPy PF is also drastically increasing from 0.8 to 3.5 ppm and its sensitivity is droping from 2.5 to 1.6 \%/ppm as it ages. In contrast, the LOD of PPy NT improved with aging from 0.15 ppm to 7 ppb and sensitivity remains relatively stable around 0.01 \%/ppm. The PPy PF is losing sensing performance with age while nanotube is gaining with increasing response and stable sensitivity across two months. The  continuous film of PPy appears more vulnerable to aging process while the NT seems comparatively robust. The low LOD and noise in baseline response in PPy NT correlates with high environmental stability and more stable morphology in the NT and is consistent with stable sensitivity and improved response over 60 days. The aging of kinetic parameter is also evaluated as a function of the time, as shown in figures \ref{fig:3}(c) and (d). Both time constants differentiate the two morphologies as they age. The time constants $\tau_1$ and $\tau_2$ of PPy NT in response and recovery is decreasing showing a faster response as it ages. While the  two time constants of PPy PF is not consistently changing as it is aging. 
\par Since the same sensor was remeasured each time, the aging effects reported combines two processes (1) the passive chemical or physical aging of the polymer and (2) cumulative conditioning from NH$_3$ exposures. This more faithfully represents the real situation of an aging sensor rather than an idealized, unexposed aging. Such unexposed aging in PPy is attributed to dopant loss and consequent charge carrier depletion \cite{varga2017}. Repeated gas exposure can lead to NH$_3$ accumulation and structural relaxation manifested by grain coarsening in PPy \cite{Amini2022} which changes the conduction network. The smaller drift in resistance in PPy NT is consistent with the residual MO and its sulfonate (SO$_3^{-1}$) acting as secondary counter-ion, partially protecting the polymer backbone from dopant loss \cite{varga2017}. In contrast, the sensor performance depends on how far NH$_3$ diffuse through bulk polymer to reach the active conduction pathway. In the PPy PF, this distance spans the continous film thickness, whereas in the PPy NT, it is limited to the thin walls of porous structure. As a result, the same bulk aging process affects the two morphologies differently, degrading the performance of PF as bulk conduction declines, while the gas accessible sites of PPy NT remain unaffected regardless of bulk conduction. The two contributions to aging cannot be separated from this dataset, but their combined effect and dependence on morphology  is the practically relevant finding for accessing long-term sensor reliability.

\section*{Conclusion}
The sensing characteristics of PPy chemiresistor in nanotube (NT) and planar film (PF) trade one performance axis for another over time. The PPy PF offers a higher response and sensitivity when freshly fabricated, but degrades on every performance metric with age. Comparatively, PPy NT despite their lower initial sensitivity has faster response time and low limit of detection (LOD), and steadily improve in terms of response, speed and detection limit over 60 days. Unlike the conventional sensors that suffer from degradation over usage, PPy NT maintain reliable sensitivity, improved LOD and NH$_3$ response as it ages without requiring frequent recalibration. 
As both samples age, NT $\tau_1$ and $\tau_2$ fall monotonically while that of PF show no consistent trend. The aging driven performance and kinetic improvement are uniquely specific to the nanotube morphology. These results support PPy nanotube as preferred choice for long-term, low maintenance NH$_3$ chemiresistor applications.
\section*{Materials and Methods}
Interdigitated electrode (IDE) substrates were fabricated by optical lithography, depositing Au/Ti (200/10 nm) on Si/SiO$_2$ wafer. Each IDE consists of 50 fingers, 250 $\mu$m in length and 2 $\mu$m in width, with a 2.5 $\mu$m gap. Template-free planar films (PF) of PPy were deposited electrochemically onto the IDE substrate under a constant potential of 0.85 V in a three electrode configuration, with the IDE as the working electrode, a standard Ag/AgCl reference electrode, and a stainless steel counter electrode. The electrolyte consisted of an aqueous mixture of 0.1 M pyrrole (Py, 99 \% purity, Thermo Scientific) and 0.1 M lithium perchlorate (LiClO$_4$, 98 \% purity, Thermo Scientific) as the counter-ion. PPy nanotubes (NT) were deposited by cyclic voltammetry onto the IDE substrate, cycling the potential from -0.4 to 1 V at a scan rate of 1 V/s, using a VersaSTAT potentiostat in the same three electrode configuration described above. The electrolyte used for nanotube growth was 0.55 M pyrrole, 0.1 M LiClO$_4$, and 12.25 mM methyl orange (MO, Thermo Scientific), with methyl orange serving as a soft templating agent to promote nanotube morphology. LiClO$_4$ and MO were used as received and pyrrole was distilled prior to use. 
Samples with different baseline resistances were obtained by varying the deposition time (PFs, [8-15 s]) or the number of CV cycles (NTs, [1-10 cycles]). The film thickness was measured by profilometry (Dektak 150, Veeco) for all samples. The PPy PF used in this study was 450 nm and the PPy NT sample was 787 nm thick.

\par Gas sensing measurements were done at room temperature in a custom-built Teflon chamber. Gas delivery was controlled by four mass flow controllers (Brooks Instrument GF series), supplying 20 \% O$_2$, NH$_3$, dry N$_2$, and N$_2$ passed through a de-ionized water bubbler to maintain 50 \% relative humidity; the NH$_3$ concentration in the chamber was set by adjusting the relative flow rates of these four lines.  The gases were mixed before entering the gas inlet of the chamber and a constant total  flow rate of 1000 cubic centimeters per minute (sccm) was maintained. Each sensor was connected to a Keysight data acquisition unit, and the resistance across the IDE was recorded using BenchVue software at a 1 s scanning interval. The chamber was purged with the humidified O$_2$/N$_2$/N$_2$ gas mixture (without NH$_3$) for 3 h to stabilize the baseline resistance and remove residual impurities from the chamber and sensing layer. NT sensors were exposed to NH$_3$ concentrations of 5–50 ppm for 75 min per exposure, with a 120 min recovery period between exposures; PF sensors were exposed to 5–43 ppm for 60 min per exposure, also with a 120 min recovery period. Devices showing anomalous blank noise behavior were excluded as outliers in the reported LOD range.

\section*{Conflicts of Interest}
There are no conflicts of interest.

\begin{acknowledgments}
	This research was supported by the Walloon Region through the BEWARE Fellowship Program (Project COPERNICAN-2410167). The authors  acknowledge VOCSens for technical support, material resources, and scientific collaboration. 
\end{acknowledgments}

 \bibliography{Ref}

\end{document}


\title{Supplementary Material  \\ \small 	Aging-Improved NH$_3$ Sensing in Polypyrrole Nanotube Chemiresistors}
\author{Arya Mohan}
\email{arya.mohan@uclouvain.be}
\affiliation{Nanoscopic Physics (NAPS), Institute of Condensed Matter and Nanosciences (IMCN), Université catholique de Louvain (UCLouvain), 1348 Louvain-la-Neuve, Belgium}

\author{Olivia Kamdom}
\affiliation{Nanoscopic Physics (NAPS), Institute of Condensed Matter and Nanosciences (IMCN), Université catholique de Louvain (UCLouvain), 1348 Louvain-la-Neuve, Belgium}

\author{Antoine Enel}
\affiliation{Institute of Information and Communication Technologies, Electronics and Applied Mathematics (ICTEAM), Université catholique de Louvain (UCLouvain), 1348 Louvain-la-Neuve, Belgium}
\affiliation{VOCSens, 6 Chemin du Cyclotron, 1348 Louvain-la-Neuve, Belgium}

\author{Beno\^{i}t Hackens}
\email{benoit.hackens@uclouvain.be}
\affiliation{Nanoscopic Physics (NAPS), Institute of Condensed Matter and Nanosciences (IMCN), Université catholique de Louvain (UCLouvain), 1348 Louvain-la-Neuve, Belgium}

\maketitle
\hspace{4em}\noindent $^{*}$arya.mohan@uclouvain.be ; \noindent$^{\dagger}$benoit.hackens@uclouvain.be \vspace{1em}	\\	
		
\begin{figure*}
	\centering
	\includegraphics[width=1\linewidth]{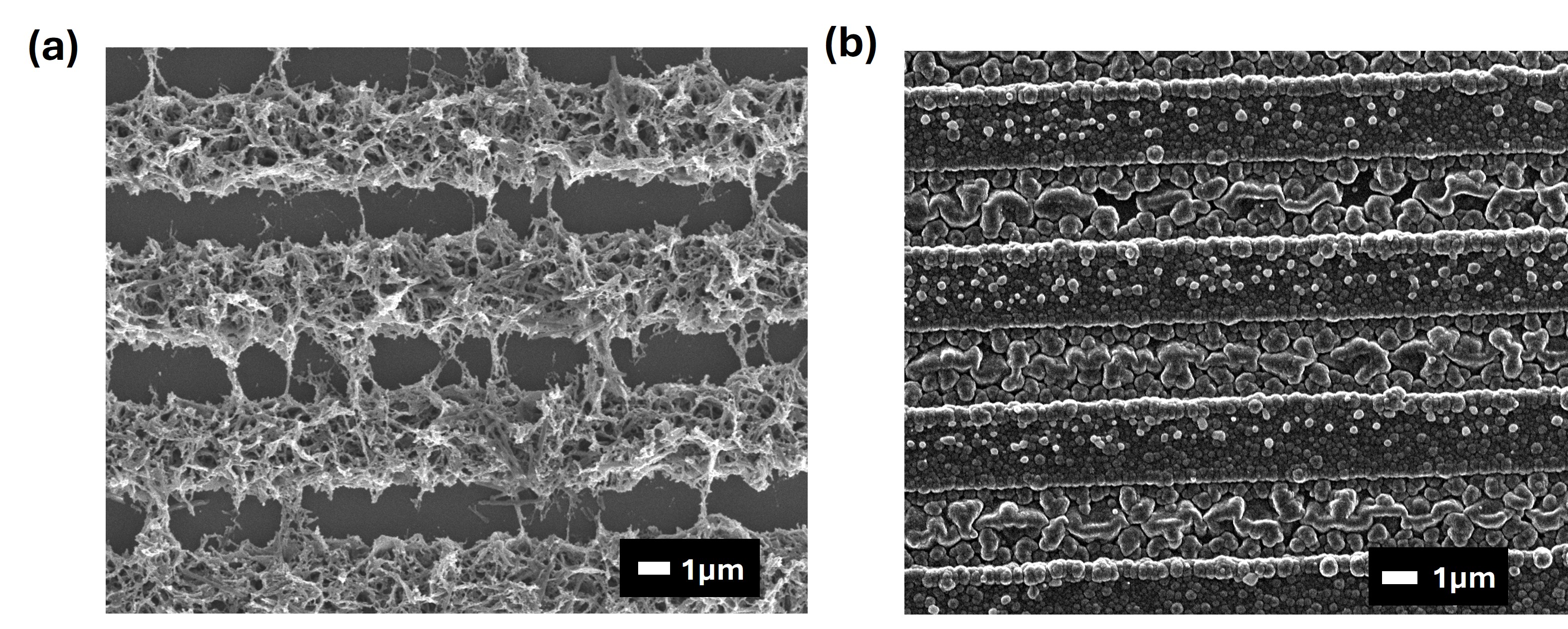}
	\caption{Scanning electron microscopy (SEM) images of (a) PPy nanotube and (b)  PPy planar film samples.}
	\label{fig:sem}
\end{figure*}

The methyl orange (MO) templated polypyrrole (PPy NTs, nanotubes) and template-free polypyrrole (PPy PFs, planar films) deposited on identical interdigitated electrodes exhibit distinct morphologies, as shown in scanning electron microscopy (JEOL JSM-7600F) in Fig. \ref{fig:sem}. Specifically, Fig. \ref{fig:sem}(a) shows the nanotube morphology of the PPy NT sensors, whereas Fig. \ref{fig:sem}(b) reveals the dense, globular morphology of the PPy PF sensors.\\
	
\newpage
	\begin{figure*}
	\centering
	\includegraphics[width=0.8\linewidth]{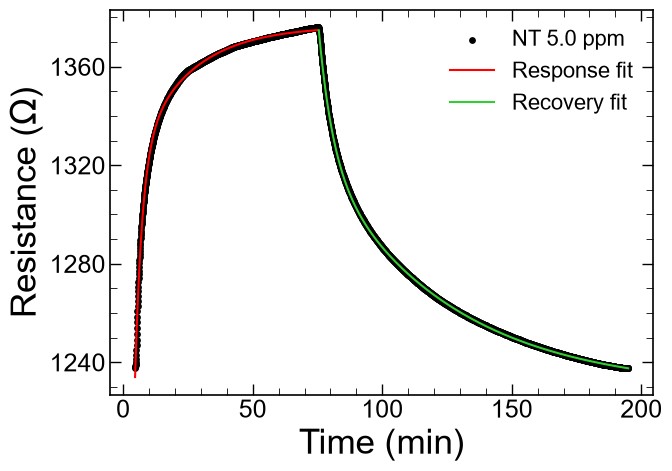}
	\caption{Double exponential fit on 5 ppm ammonia (NH$_3$) response and recovery of PPy nanotube (NT).}
	\label{fig:S1}
\end{figure*}
The response and recovery curves of both PPy NT and PPy PF were fitted to a double-exponential model (eqns. \ref{eq:S1} \& \ref{eq:S2}), with a representative fit for PPy NT exposed to $5\text{ ppm}$ ammonia ($\text{NH}_3$) shown in Fig. \ref{fig:S1}.
The resistance curves during exposure and recovery were fitted to double-exponential functions:\\

\noindent{Response:}
\begin{equation}
	R(t) = R_{\text{final}} - (R_{\text{final}} - R_0)\left[A\exp\left(-\frac{t}{\tau_1}\right) + (1-A)\exp\left(-\frac{t}{\tau_2}\right)\right]
	\label{eq:S1}
\end{equation}

\noindent{Recovery:}
\begin{equation}
	R(t) = R_{\text{final}} + (R_0 - R_{\text{final}})\left[A\exp\left(-\frac{t}{\tau_1}\right) + (1-A)\exp\left(-\frac{t}{\tau_2}\right)\right]
	\label{eq:S2}
\end{equation}

\noindent where $R_0$ is the initial resistance immediately  before the exposure or recovery, $R_{\text{final}}$ is the resistance at the end of the exposure or recovery, $t$ is the elapsed time since the onset of exposure or recovery, and $\tau_1$ and $\tau_2$ are the fast and slow time constants, respectively. The parameter $A_1$ represents the relative amplitude weight of the fast process ($0 \le A_1 \le 1$), with $(1-A_1)$ denoting the amplitude of the slow process. Each curve was fitted independently to Eq.~\ref{eq:S1} (response) or Eq.~\ref{eq:S2} (recovery) for every concentration. The extracted time constants and relative amplitudes as a function of $\text{NH}_3$ concentration are summarized in Tables~\ref{nanotube} and~\ref{bulk}, while their evolution over aging is in Table~\ref{aging}.

	\begin{table}
		\centering
		\renewcommand{\arraystretch}{1.5}
		\caption{Response and recovery time constants ($\tau_1$ \& $\tau_2$) and amplitude (A$_1$) of PPy nanotube.}
		\begin{tabular}{|c|c|c|c|c|c|c|}
			\hline
			\multirow{2}{*}{Concentration (ppm)} & \multicolumn{3}{c}{Response} & \multicolumn{3}{|c|}{Recovery} \\
			\cline{2-7}
			& $\tau_1$(min) & $\tau_2$(min) & A$_1$ & $\tau_1$(min) & $\tau_2$(min) & A$_1$ \\
			\hline
			5 & 3.24 & 30.68 & 0.86 & 4.67 & 36.96 & 0.35 \\
			\hline
			10 & 2.19 & 43.78 & 0.84 & 5.24 & 60.81 & 0.52 \\
			\hline
			20 & 1.73 & 24.94 & 0.87 & 3.06 & 28.21 & 0.52 \\
			\hline
			50 & 1.26 & 36.54 & 0.79 & 3.59 & 53.78 & 0.59 \\
			\hline
		\end{tabular}
		\label{nanotube}
	\end{table}
	
	\begin{table}
		\centering
		\renewcommand{\arraystretch}{1.5}
		\caption{Response and recovery time constants ($\tau_1$ \& $\tau_2$) and amplitude (A$_1$) of PPy planar film.}
		\begin{tabular}{|c|c|c|c|c|c|c|}
			\hline
			\multirow{2}{*}{Concentration (ppm)} & \multicolumn{3}{c}{Response} & \multicolumn{3}{|c|}{Recovery} \\
			\cline{2-7}
			& $\tau_1$(min) & $\tau_2$(min) & A$_1$ & $\tau_1$(min) & $\tau_2$(min) & A$_1$ \\
			\hline
			5 & 3.40 & 40.22 & 0.76 & 7.63 & 43.60 & 0.41 \\
			\hline
			10 & 3.15 & 31.44 & 0.79 & 6.25 & 50.57 & 0.37 \\
			\hline
			22 & 2.62 & 43.54 & 0.76 & 4.47 & 42.36 & 0.48 \\
			\hline
			43 & 2.50 & 43.41 & 0.51 & 4.32 & 43.25 & 0.42 \\
			\hline
		\end{tabular}
		\label{bulk}
	\end{table}

	\begin{table}
		\centering
		\renewcommand{\arraystretch}{1.5}
		\caption{Response and recovery time constants ($\tau_1$ \& $\tau_2$) and amplitude (A$_1$) as a function of aging time.}
		\begin{tabular}{|c|c|c|c|c|c|c|c|}
			\hline
			\multirow{2}{*}{Samples} & \multirow{2}{*}{Aging time (Days)} & \multicolumn{3}{c}{Response} & \multicolumn{3}{|c|}{Recovery} \\
			\cline{3-8}
			& & $\tau_1$(min) & $\tau_2$(min) & A$_1$ & $\tau_1$(min) & $\tau_2$(min) & A$_1$ \\
			\hline
			\multirow{3}{*}{Bulk} & 0 & 3.15 & 31.44 & 0.79 & 6.25 & 50.57 & 0.37 \\
			\cline{2-8}
			& 30 & 3.54 & 21.83 & 0.83 & 8.26 & 54.26 & 0.32 \\
			\cline{2-8}
			& 45 & 2.71 & 26.05 & 0.89 & 6.05 & 63.86 & 0.36 \\
			\hline
			\multirow{3}{*}{Nanotube} & 0 & 2.19 & 43.78 & 0.84 & 5.24 & 60.81 & 0.52 \\
			\cline{2-8}
			& 30 & 2.11 & 23.00 & 0.87 & 4.46 & 46.17 & 0.43 \\
			\cline{2-8}
			& 60 & 1.62 & 13.77 & 0.87 & 4.00 & 37.51 & 0.42 \\
			\hline
		\end{tabular}
		\label{aging}
	\end{table}